\documentclass[twocolumn,showpacs,preprintnumbers,amsmath,amssymb]{revtex4}

\usepackage{graphicx}
\usepackage{dcolumn}
\usepackage{bm}

\begin{document}

\preprint{ICN-UNAM/123}

\title{Conformal Anisotropic Mechanics}

\author{Juan M. Romero}
\email{jromero@correo.cua.uam.mx}

\affiliation{%
Departamento de Matem\'aticas Aplicadas y Sistemas\\
Universidad Aut\'onoma Metropolitana-Cuajimalpa, M\'exico 11950 DF,
M\'exico}%
\author{Vladimir Cuesta}
 \email{vladimir.cuesta@nucleares.unam.mx}
 \author{J. Antonio Garcia}%
 \email{garcia@nucleares.unam.mx}
 \author{J. David Vergara}%
 \email{vergara@nucleares.unam.mx}
\affiliation{
Instituto de Ciencias Nucleares, Universidad Nacional Aut\'onoma de M\'exico,\\
Apartado Postal 70-543, M\'exico 04510 DF, M\'exico
}%

\date{\today}

\begin{abstract}
In this paper we implement scale
anisotropic  transformations in the space-time in classical mechanics. The
resulting system is consistent with the dispersion relation of gravity at a
Lifshitz point recently considered in \cite{horava:gnus}.
Also, we show that our model is a generalization of
the conformal mechanics of Alfaro, Fubini and Furlan. For arbitrary
$z$ we construct the dynamical symmetries that correspond to the Schroedinger algebra. Furthermore, we obtain the Boltzman distribution for a gas of free particles compatible
with anisotropic scaling transformations and compare our result with the corresponding thermodynamics of the recent anisotropic black branes proposed in \cite{amanda:gnus}.
\end{abstract}

\pacs{Valid PACS appear here}
\maketitle

\section{\label{sec:level1}Introduction}

Scaling space-time symmetries are very useful tools to study non
linear physical systems and critical phenomena
\cite{condensada:gnus,Cardy:gnus}. At the level of classical
mechanics an interesting example is the conformal mechanics
\cite{fubini:gnus} that is relevant for several physical systems
from molecular physics to black holes
\cite{strominger:gnus,anomalia:gnus}. Recently a new class of
anisotropic scaling transformations has been considered in the
context of critical phenomena \cite{Henkel:gnus}, string theory
\cite{Son, barbon:gnus,amanda:gnus} and quantum gravity
\cite{horava:gnus}.
\begin{eqnarray}
t\to b^{z}t,\qquad \vec x\to b \vec x. \label{eq:escala}
\end{eqnarray}
where $z$ play the role of a dynamical critical exponent. In
particular the result of Horava \cite{horava:gnus} is quite
interesting since it produces a theory that reduces to general
relativity at large scales and may provide a candidate for a UV
completion of general relativity. In principle this model for
Gravity is renormalizable in the sense that the effective coupling
constant is dimensionless in the UV limit. A fundamental property of
these gravity models is that the scaling laws (\ref{eq:escala}) are
not compatible with the usual relativity. In the IR these systems
approach the usual General Relativity with local Lorentz invariance
but in the UV the formulation admit generalized dispersion relations
of the form
\begin{eqnarray}
P_{0}^{2}-G(\vec P^{2} )^{z}=0, \qquad G={\rm const.}
\label{eq:horava}
\end{eqnarray}
Where $z$ plays the role of a dynamical exponent. As the dispersion
relations used by these models are quadratic in $P_0$ while the
spatial momentum scale as $z$ the models are in principle
renormalizable
 by power counting arguments at least for $z=3$. It is claimed also that these models are
 ghost free.  On the other hand due to the non
relativistic character of the Horava gravity, this theory must be
considered as an effective version of a more fundamental theory,
compatible with the symmetries that we observe in our universe.
The consideration of these models could be useful to
understand the quantum effects of gravity.

In order to gain some understanding of the content of generalized dispersion relations of
the form (\ref{eq:horava}) we will construct a reparametrization invariant classical mechanics compatible with the
scaling laws (\ref{eq:escala}). In particular this implementation of the
scaling transformations (\ref{eq:escala}) will lead to
dispersion relations of the form (\ref{eq:horava}). Another interesting
characteristic of our implementation is that it contains interesting systems in the limits  $z=2$  where the
conformal mechanics of Fubini, et al \cite{fubini:gnus} is recovered,
$z=1$ where we recover the standard relativistic particle and
$z \to \infty$ where we obtain the Euclidian relativistic particle.

 We also study the symmetries of the anisotropic mechanics and compare our results with the Schroedinger algebra for any value of $z$.
 Finally, by
analyzing the thermodynamic properties of a free gas compatible with the scaling transformation (\ref{eq:escala}) we will obtain the same thermodynamic relations
for black branes recently proposed in \cite{amanda:gnus}.

The organization of the paper is as follows: Section 2 concerns with
the action of the system. In Sec. 3 we study the canonical formalism
of the system and his gauge symmetries. Sec. 4 focuses on the
equations of motion. The global symmetries and the Schroedinger
algebra are analyzed in section 5. In Sec. 6 we consider the
thermodynamic properties of the system and we conclude in section 7.

\section{The action principle}

Scale transformation can be realized in non-relativistic theories in
different ways. From one hand  we can start from an action in curved
space where some of its isometries are the scale transformations.
Another way to implement scale transformations is to start from an
action in flat space but constructed in such way that the invariance
under scale transformations is manifest. In this work will take this
second point of view. Let us consider a $d+1-$dimensional
space-time, with an action principle given by
\begin{eqnarray}
\label{anMech}
&&S=\int d\tau L=
\int d\tau \left(\frac{m}{2} \frac{\left( \dot  x^{2}\right)^{\frac{z}{2(z-1)}}}
{\left(\dot t\right)^{\frac{1}{z-1}}}  - V\left( \vec x\right)\dot t \right ), \label{eq:accion} \\
&& \quad \dot t=\frac{dt}{d\tau},\quad  \dot x_{i}=\frac{d
x_{i}}{d\tau},\quad  \dot x^{2}= \dot x_{i}\dot x^{i},\quad  i=1,
2,\cdots , d.\nonumber
\end{eqnarray}
This action is invariant under reparametrizations of $\tau$, i.e.
under the transformations $\tau \to \tau=\tau(f),$ if we require
that $t$ and the $x_i$ are scalars under reparametrizations.
Furthermore, if we consider that the potential  satisfies $V(b\vec
x)=b^{-z}V(\vec x),$ for example,
\begin{eqnarray}
V(\vec x)=\frac{a}{|\vec x|^{z}},\qquad a={\rm cte}
\label{eq:potencial}
\end{eqnarray}
the action (\ref{eq:accion}) is invariant under the scaling laws
Eq.(\ref{eq:escala}). Note that, when $z=2$ and $\tau =t$ we obtain the action for a
non-relativistic particle, furthermore taking into account (\ref{eq:potencial}), we obtain
\begin{eqnarray}
S=\int dt \left(\frac{m}{2} \dot  x^{2}  - \frac{a}{|\vec x|^{2}}
\right ),
\end{eqnarray}
that corresponds to the action of the conformal mechanics
\cite{fubini:gnus}. In the case of arbitrary $z$ and for a gauge condition $\tau =t$ we get
\begin{eqnarray}
S=\int dt \left(\frac{m}{2} \left(\dot
x^{2}\right)^{\frac{z}{2(z-1)}}  - \frac{a}{|\vec x|^{z}}  \right ),
\label{eq:mconforme}
\end{eqnarray}
that again is invariant under the scaling laws
(\ref{eq:escala}). In this sense Eq.(\ref{eq:accion}) represents a generalized conformal
mechanics.\\

Another interesting property of the action (\ref{eq:accion}) appears in the limit  $z\to \infty$. In this
case we obtain
\begin{eqnarray}
S=\int d\tau \left(\frac{m}{2} \sqrt{\dot x^{2}}  - V\left( \vec
x\right)\dot t \right ),
\end{eqnarray}
this action is invariant under reparametrizations, and leaving
$V(\vec x)=0$  we obtain the action for an Euclidean relativistic
particle
\begin{eqnarray}
S=\int d\tau \frac{m}{2} \sqrt{\dot x^{2}}. \label{eq:euclidiana}
\end{eqnarray}
Another interesting limit is in the case $z=1$. Even though this limit is not well defined for the action (\ref{anMech}) it can be implemented using an equivalent form of the action   (\ref{anMech})
\begin{equation}\label{act-e}
S=\int d\tau \frac{m}{2(z-1)}\left(\frac{(\dot x^2)^{\frac{z}{2}}}{e^{z-2}\dot t}+(z-2)e\right).
    \end{equation}
Here, we have introduced an einbein $e$ in a similar way as the standard trick to remove the
square root in the action of the free relativistic particle. Using this action (\ref{act-e}) we can now
take the limit $z\to 1$ if at the same time we take the limit $m\to 0$. Assuming that $m\to 0$ at the same rate as $z-1$ in such way that $\frac{m}{z-1}\to 1$ we obtain an action that is very similar to the action of the massless relativistic free particle.
We conclude the the limit $z\to 1$ corresponds to the massless relativistic particle that is invariant under the full group of relativistic conformal transformations.

\section{Canonical formalism}

Let us consider the canonical formalism of the system
(\ref{eq:accion}). The canonical momenta are
\begin{eqnarray}
p_{i}\!\!&=&\!\!\frac{\partial L}{\partial \dot x^{i} }=\frac{mz}{2(z-1)}
\frac{(\dot x^{2})^{\frac{2-z}{2(z-1)}}}{ \left(\dot t\right)^{\frac{1}{z-1}}} \dot x_{i},\label{eq:momento}\\
p_{t}\!\!&=&\!\!\frac{\partial L}{\partial \dot t }=\!\!
-\left(\frac{m}{2(z-1)}\frac{(\dot x^{2})^{\frac{z}{2(z-1)}}}{
\left(\dot t\right)^{\frac{z}{z-1}}}+V(\vec x)\right)\!\!=\!\!-H.
\end{eqnarray}
and as expected (by the reparametrization invariance) the canonical Hamiltonian vanishes
\begin{eqnarray}
H_{c}=p_{i}\dot x^{i}+p_{t}\dot t-L=0.
\end{eqnarray}
From (\ref{eq:momento}) we get
\begin{eqnarray}
p^{2}=p_{i}p^{i}&=&\left( \frac{mz}{2(z-1)} \frac{(\dot
x^{2})^{\frac{1}{2(z-1)}}}{ \left(\dot t\right)^{\frac{1}{z-1}}}
\right)^{2},
\end{eqnarray}
that implies in turn
\begin{eqnarray}\label{cons2}
\phi=p_{t}+\frac{m}{2(z-1)}  \left( \frac{mz}{2(z-1)}
\right)^{-z}\left(p^{2}\right)^{\frac{z}{2}}+V(\vec x)\approx 0.
\label{eq:constriccion}
\end{eqnarray}
If we enforce the constraint, and considering the case when $V(\vec x)=0$ we obtain
\begin{eqnarray}
\label{anSchreq}
p_{t}=-\frac{m}{2(z-1)}  \left( \frac{mz}{2(z-1)}
\right)^{-z}\left(p^{2}\right)^{\frac{z}{2}},
\end{eqnarray}
which can be rewritten in the form
\begin{eqnarray}
\left(p_{t}\right)^{2}-G \left(p^{2}\right)^{z}=0,\qquad
G=\frac{1}{z^{2z}} \left(\frac{m}{2(z-1)}\right)^{2(1-z)}.
\end{eqnarray}
In this way, we get a dispersion relation that is very similar to the Horava's gravity model
Eq. (\ref{eq:horava}).\\

Notice that by taking the limits $z\to 1$ and $m\to 0$  in (\ref{cons2}) we get the constraint
\begin{equation}\label{cons3}
\phi_1= p_t + (p^2)^{\frac{z}{2}}+V(\overrightarrow{x}),
\end{equation}
that is consistent with the canonical formalism of the action (\ref{act-e}) and corresponds to a
massless relativistic particle with only positive energy.

The system has one first class constraint (\ref{eq:constriccion})
and following the standard Dirac's method  \cite{dirac:gnus},  the
``wave equation"
\begin{eqnarray}
\hat \phi|\psi> =\left(\hat p_{t}+\sqrt{G}\left(\hat
p^{2}\right)^{\frac{z}{2}} +V(\vec x)\right)|\psi> =0.
\end{eqnarray}
must be implemented on the Hilbert space at quantum level to obtain the physical states.
In other words, the "Schroedinger equation" corresponding to the anisotropic mechanics (\ref{anMech}) is
\begin{eqnarray}
-i\hbar\frac{\partial \psi}{\partial t}=\left(\sqrt{G}
\left(-\hbar^{2}\nabla ^{2}\right)^{\frac{z}{2}} +V(\vec x)\right)
\psi.
\end{eqnarray}

\subsection*{Gauge of freedom}

Because of the action (\ref{eq:accion}) is invariant under
reparametrizations, there is a local symmetry. The extended
canonical action for the system is \cite{henneaux:gnus}
\begin{eqnarray}
S=\int d\tau \left(\dot t p_{t}+\dot x^{i} p_{i}-\lambda \phi\right)
\end{eqnarray}
with $\lambda$ a Lagrange multiplier. \\

If we require that $\epsilon$ is a function of $\tau$ the extended action is invariant under the gauge transformations
\cite{henneaux:gnus}
\begin{eqnarray}
\delta x_{i\epsilon}&=& \{x_{i},\epsilon \phi\}=\epsilon \sqrt{G}z(p^2)^{\frac{z-2}{2}} p_{i},\nonumber \\
\delta p_{i\epsilon}&=& \{p_{i},\epsilon \phi\}=-\epsilon \frac{\partial V}{\partial x_{i}},\nonumber\\
\delta t_{\epsilon}&=& \epsilon \{t,\phi\}=\epsilon ,\nonumber \\
\delta p_{t\epsilon}&=& 0,\nonumber\\
\delta \lambda_{\epsilon}&=& \frac{\partial \epsilon}{\partial \tau}.\nonumber
\end{eqnarray}
Here $\{,\}$ are the Poisson brackets. The above transformations, are of the usual type for the parametrized particle. However, the rule for the transformation corresponding to the spatial coordinates seems to change drastically. But if we use the definition of the momenta (\ref{eq:momento}) we will get exactly the usual transformation $\delta x_{i\epsilon}=\epsilon \frac{dx^i}{dt}$ recovering the full diffeomorphism transformations associated with the reparametrization invariance. This result can be constrasted with the same result obtained by Horava's in his gravity model \cite{horava:gnus}.\\

In this way we saw that the action Eq.(\ref{eq:accion}), has very similar properties to the Horava's gravity namely the same scaling properties and the same invariance under reparametrizations. In this sense, we see that our toy model has the same transformation properties than the Horava's gravity just as the relativistic particle has the same transformation properties as General Relativity \cite{teitelboim:gnus}.

\section{Equations of motion}

From the variation of the action (\ref{eq:accion}), and taking into account the definitions of the momenta $p_{t}$ and $p_{i},$ it follows that
\begin{eqnarray}\label{eq:variacion}
\delta S&=&\int d\tau\left(\frac{d}{d\tau}\left( p_{i} \delta
x_{i}+p_{t}\delta t\right) \right. \\&& \left.- \delta x_{i} \left( \frac{d
p_{i}}{d\tau} + \frac{\partial V}{\partial x_{i}}\right) +\delta t
\frac{d p_{t} }{d\tau}\right), \nonumber
\end{eqnarray}
where the Euler-Lagrange equations are
\begin{eqnarray}
& & \frac{d p_{i}}{d\tau}+ \frac{\partial V}{\partial x_{i}}=0,
\label{eq:movimiento1}\\
& & \frac{d}{d\tau}p_{t}=0. \label{eq:movimiento2}
\end{eqnarray}
Notice that for $z=2$ and $\tau =t$ the equation (\ref{eq:movimiento1}) we obtain the usual Newton's second law, and that (\ref{eq:movimiento2}) correspond to the energy conservation.\\

In order to simplify some computations we will use the gauge condition
$\tau=t.$ This is a good gauge condition in the sense that $\{\phi,\tau-t\}\not =0.$ Under the above assumption and  using the definition  $p_{i}$ we get the equations of motion
\begin{eqnarray}
\frac{d}{dt}\left( \left(\frac{mz}{2(z-1)}\right)\left(\dot
x^{2}\right)^{\frac{2-z}{2(z-1)}}\dot x^{i}\right)+ \frac{\partial
V}{\partial x_{i}}=0,
\end{eqnarray}
We thus get
\begin{eqnarray}
g_{ij} \ddot x_{j}&=&-\left(\dot x^{2}\right)^{\frac{z-2}{2(z-1)}}
\frac{2(z-1)}{mz} \frac{\partial V}{\partial x_{i}}, \nonumber\\
g_{ij}&=&\left( \delta_{ij} +\frac{2-z}{z-1} \frac{\dot x_{i}\dot
x_{j}}{\dot x^{2}}\right).
\end{eqnarray}
The matrix $g_{ij}$ can be considered as a metric that depends on the velocities.
We see that this metric is homogeneous of degree zero in the velocities and in consequence is a Finsler type metric \cite{finsler:gnus}. This metric has three critical points:
\begin{eqnarray}
z=1,\qquad z=2,\qquad z=\infty.
\end{eqnarray}
For $z\not = \infty,$ we obtain the inverse metric
\begin{eqnarray}
  g^{ij}=\left( \delta^{ij} -(2-z) \frac{\dot x^{i}\dot x^{j}}{\dot x^{2}}\right),
\end{eqnarray}
it follows that
\begin{eqnarray}
\ddot x_{i}= \frac{2(1-z)}{mz} \left(\dot
x^{2}\right)^{\frac{z-2}{2(z-1)}}
 \left( \delta_{ij} +(z-2) \frac{\dot x_{i}\dot x_{j}}{\dot x^{2}}\right)  \frac{\partial V}{\partial x_{j}}.
\end{eqnarray}
The limit $z=\infty$  is particularly interesting, since in this
case the matrix $g_{ij}$ is not invertible. This means that the
gauge condition $\tau=t$ is not valid.  Because of in this limit we get
the action (\ref{eq:euclidiana}) and in this case is not more valid the constraint (\ref{eq:constriccion}).\\

\section{Dynamical Symmetries of the Schroedinger equation for any $z$}

In the same sense as the Galilei group  exhaust all the symmetries
of the nonrelativistic free particle in $d$ dimensions the
Schroedinger Algebra is the algebra of symmetries of the
Schroedinger equation with $V=0$ and can be considered as the non
relativistic limit of the Conformal symmetry algebra. Another
interesting Galilean Conformal Algebra can also be constructed as
the non relativistic contraction of the full Conformal Relativistic
Algebra \cite{Gopa}. The generators of the Schroedinger algebra
include temporal translations $H$, spatial translations $P^i$,
rotations $J^{ij}$, Galilean boost $K^i$, dilatation $D$ (where time
and space can dilate with different factors), special conformal
transformation $C$, and the mass operator $M$.The nonzero
commutators of this algebra are the Galilean algebra
\begin{eqnarray}
&&\left\{J^{ij},J^{kl}\right\}=(\delta^{ik}J^{jl}+\delta^{jl}J^{ik}-\delta^{il}J^{jk}-\delta^{jk}J^{il}),\nonumber\\
&&\left\{J^{ij},P^k\right\}=(\delta^{ik}P^j-\delta^{jk}P^i), \nonumber\\
&&\left\{J^{ij},K^k\right\}=(\delta^{ik}K^j-\delta^{jk}K^i), \nonumber\\
 && \left\{H,K^i\right\}=P^i\nonumber\\
 \label{Gali}
 &&\left\{P^i,K^j\right\}=-\delta^{ij}M,
\end{eqnarray}
plus Dilatations
\begin{equation}
\left\{D,P^i\right\}=P^i,\quad \left\{D,K^i\right\}=K^i, \quad
\left\{D,H\right\}=-2H
\end{equation}
and Special Conformal transformations
\begin{equation}
\left\{D,C\right\}=-2C,\quad \left\{H,C\right\}=D, \quad
\left\{C,P^i\right\}= K^i.
\end{equation}
It is interesting to notice that $H,C,D$ close themselves in an
$SL(2,\mathbb{R})$ subalgebra. Thus the Schroedinger algebra is the
Galilean algebra  plus  dilatation and Special Conformal
transformations \cite{Hagen-Niederer}. $M$ is the center of the
algebra.

It is also worth noticing that the Schroedinger algebra in $d$ dimensions, $Schr(d)$,  can be embedded
into the relativistic conformal algebra in $d+2$ space-time dimensions $O(d+2,2)$.  This fact is central
in the recent literature about the geometric content of the new AdS/NRCFT dualities (for details see \cite{Son}).

We are interested in the construction of the explicit generator of
the relative Poisson-Lie Schroedinger algebra for any $z$. These
will be the symmetries of the corresponding anisotropic Schroedinger
equation (\ref{anSchreq}) associated with the non-relativistic
anisotropic classical mechanics (\ref{anMech}).

We will denote the generalization of the Schroedinger symmetry algebra for
any $z$ as $Schr_z(d)$ in $d$ dimensions. The algebra is given by the Galilean algebra (\ref{Gali}) plus
\begin{eqnarray}
&&\left\{D,P^i\right\}=P^i,\quad \left\{D,K^i\right\}=(1-z)K^i, \nonumber\\
\label{SchrZ}
 &&\left\{D,M\right\}=(2-z)M,\quad \left\{D,H\right\}=-zH
\end{eqnarray}
Notice that $M$ is not playing the role of the center anymore unless
$z=2$ and that $C$, the generator of the Special Conformal
transformations is not in the algebra \cite{BMcG}. We are not aware
of any explicit realization of this algebra in phase space for
arbitrary dynamical exponent $z$. For $z=2$ the phase space
realization reads
\begin{eqnarray}
&&J^{ij}=x^ip^j-x^jp^i, \quad H=-p_t,\quad P^i=p^i,\nonumber \\
&& K^i=-Mx^i+tp^i\nonumber\\ \nonumber &&C=t^2p_t+t x^i p^i-\frac12
M x^2,\quad D=2tp_t+ x^ip^i
\end{eqnarray}
Inspired by this construction and the embedding of the Schroedinger
algebra in $d$ dimensions into the full relativistic Conformal
algebra in $d+2$ \cite{Son} we will display a set of dynamical
symmetries of the Schroedinger equation for any $z$.

The crux of the argument is to allow $M$ to depend on the magnitude of $p^i$ squared $M(p^2)$. We will
make the anzats that $M$ is a homogeneus function of degree $\alpha$ in  $p^i$
\begin{equation}
\frac{\partial M}{\partial p^i}p^i=2\alpha M,\qquad
M(p^2)=A(p^2)^\alpha,
\end{equation}
where $A$ is a constant. Then taking as our Schroedinger equation
(\ref{cons2}) with $V(\overrightarrow{x})=0$ and for any $z$ we have
\begin{equation}
\label{Schr}
S=p_t+H=p_t+\frac{p^2}{2M(p^2)}
\end{equation}
we will ask for the phase space quantities that commutes (in Poisson sense) with the equation (\ref{Schr}), i.e.,
\begin{equation}
\label{dynCond} \left\{{\cal O},S\right\}=0
\end{equation}
over $S=0$. Any such phase space quantity will be a dynamical symmetry of the anisotropic
Schroedinger equation (\ref{anSchreq}). $\alpha$ is fixed in terms of the dynamical exponent $z$ by
relating our definition (\ref{Schr}) with the first class constraint (\ref{anSchreq})
\begin{equation}\label{alpha}
\alpha=\frac{2-z}{2},\qquad A=\frac{1}{2\sqrt{G}}
\end{equation}
$J^{ij}, P^i, H$ are obvious symmetries of the anisotropic
Schroedinger equation. To see what are the analogs of $D,C$ and
$K^i$ let us start with $D$ as defined for the $z=2$ case but with
an anisotropic reescaling between space and time
$$D=tp_t+\beta x^ip_i$$
By taking the commutator with $S$ we have
$$\left\{D,S\right\}=p_t+\frac{\beta(1-\alpha)p^2}{M(p^2)}$$
Then
$$\beta=\frac{1}{2(1-\alpha)}$$
Using the same idea with the Special Conformal generator we propose
$$
C=t^2p_t+\gamma tx^ip_i+\sigma M(p^2) x^2
$$
where $\gamma,\sigma$ are constants to be determined from the requirement (\ref{dynCond}). The solutions are
$$
\gamma=-\frac{1}{(1-\alpha)},\qquad \sigma=-\frac{2}{(1-\alpha)^2}
$$
For the Galilean boost we have
$$ K^i=tp^i+\rho M(p^2) x^i$$
and the solution for the constant $\rho$ using the condition (\ref{dynCond}) is
$$
\rho=-\frac{1}{(1-\alpha)}
$$
So we have  the non trivial dynamical symmetries of the anisotropic Schroedinger equation generated by
\begin{eqnarray}
\label{DZ}
&&D=z tp_t+  x^ip_i,\\
\label{CZ}
&& C=\frac{z^2}{4}t^2p_t+ \frac{z}{2}t x^ip_i -\frac{1}{2}M(p^2) x^2,
\end{eqnarray}
that are dilatation and Special Conformal transformations and $d$ Galilean boosts
\begin{equation}
\label{GZ}
 K^i=\frac{z}{2}tp^i-M(p^2) x^i.
 \end{equation}
A similar result was obtained previously in \cite{Henkel:gnus}  for a differential representation
of the algebra (\ref{SchrZ}) using the concept of fractional derivatives.

Unfortunately the dynamical symmetries given by (\ref{DZ}),
(\ref{CZ}) and (\ref{GZ}) plus $J^{ij},P^i,H,M$ do not close in
$Schr_z(d)$ (\ref{SchrZ}). Nevertheless a subalgebra formed by
$H,C,D$ indeed close into an $SL(2,\mathbb{R})$ sector of the full
algebra,
$$
 \left\{D,H\right\}=-zH, \qquad \left\{D,C\right\}=-z C,\qquad \left\{H,C\right\}=\frac{z}{2}D,
$$
 To these generator we can add the angular momentum $J^{ij}$ that commutes with them. From the other hand the anisotropic system (\ref{anMech}) is also invariant under the Poincar\'e symmetry alone.

\section{Thermodynamic Properties}

In the following we shall sketch some of the thermodynamic properties of the system. We will consider only the case without potential. In this case the energy of our model is
\begin{eqnarray}
H=\sqrt{G}p^{z}.
\end{eqnarray}
Let us denote by $ {\bf V}$ the volume and $\beta =\frac{1}{kT},$ with $T$ the temperature and $k$ the Boltzmann constant, the canonical partition function for the system in a space of $d$ dimensions is
\begin{eqnarray}
Z&=&\frac{\bf V}{\hbar^{N}} \int d\vec p e^{-\beta\sqrt{G}p^{z}}=\frac{\bf V}{\hbar^{N}} \int d\Omega_{d} dp p^{d-1} e^{-\beta\sqrt{G}p^{z}}\nonumber\\
&=&\frac{{\bf V}\Omega_{d}}{\hbar^{N}} \int_{0}^{\infty}  dp p^{d-1}
e^{-\beta\sqrt{G}p^{z}}.
\end{eqnarray}
Using $u=\beta\sqrt{G}p^{z}$, and the definition of the Gamma function we obtain
\begin{eqnarray}
Z=\frac{{\bf V}\Omega_{d}}{z\hbar^{N}\left(\beta
\sqrt{G}\right)^{\frac{d}{z}} } \Gamma\left(\frac{d}{z}\right).
\end{eqnarray}
In this way, if $N$ is the number of particles, the Helmholtz free energy is given by
\begin{eqnarray}
F=-kTN\ln Z=- kTN\ln\left(\frac{{\bf
V}\Omega_{d}}{z\hbar^{N}\left(\beta \sqrt{G}\right)^{\frac{d}{z}}
}\Gamma\left(\frac{d}{z}\right)\right).
\end{eqnarray}
From this expression we may now obtain all the thermodynamic properties of the system, in particular the internal energy is
\begin{eqnarray}
U=NkT^{2}\frac{\partial \ln Z}{\partial T}=\frac{d}{z}NkT.
\end{eqnarray}
where $N$ is the total number of particles in the gas.
Substituting the number of particles $N$ in terms of the partition function ($Z=N$ in our normalization) the internal energy can be written as
\begin{eqnarray}\label{fene}
U=\frac{{\bf V}\Omega_{d}\Gamma\left(\frac{d}{z}\right) }
{z\hbar^{N}\left(\sqrt{G}\right)^{\frac{d}{z}} }
\frac{d}{z}\left(kT\right)^{\frac{d+z}{z}}.
\end{eqnarray}
Whereas the entropy is
\begin{eqnarray}\label{entro}
S=\int dT\frac{1}{T}\left(\frac{\partial U}{\partial T}\right)_{{\bf
V}} =\frac{{\bf V}\Omega_{d}\Gamma\left(\frac{d}{z}\right)
\left(k\right)^{\frac{d+z}{z}}}
{z\hbar^{N}\left(\sqrt{G}\right)^{\frac{d}{z}} }
\frac{d}{z}\left(T\right)^{\frac{d}{z}}\frac{d+z}{d}.
\end{eqnarray}
We conclude from (\ref{fene}) and (\ref{entro}), that the relationship between energy and entropy in our system is given by
\begin{eqnarray}
U=\frac{ST d}{d+z}.
\end{eqnarray}
This relationship is exactly the obtained in the case of black branes \cite{amanda:gnus}.

\section{Conclusions}

In this work we have introduced an action invariant under
anisotropic transformations in space and time. This anisotropic
mechanical system is consistent with the non-relativistic Horava's
dispersion relation. Furthermore, it was shown that for our particle
model our system has the same local symmetries. Another interesting
point is that our system corresponds to the Conformal Mechanics of
\cite{fubini:gnus} for $z=2$ and is a generalization of this system
for arbitrary $z$. Also our system includes  in the limit
$z\to\infty$ a Euclidean relativistic particle and in the limit
$z\to 1$ a massless relativistic particle. From the equations of
motion we show that naturally appears a Finsler type metric that
could implies that the Horava's gravity can be related to  Finsler
geometry. Also, we studied the dynamical symmetries of our model and
we found all dynamical symmetries of the anisotropic Schroedinger
equation for arbitrary $z$ that correspond to the generators of the
Schroedinger algebra $Schr_z(d)$. We show that the full Schroedinger
algebra constructed from our generators does not close. However we
found that a subalgebra $SL(2,\mathbb{R})$ indeed close. An
interesting point is that the explicit realization of the generators
is not linear and are analogous to the realization given in
\cite{Henkel:gnus}. As a final point, we remark that the
thermodynamic properties of our model reproduces the same
thermodynamic properties of the recent anisotropic black branes.

\begin{acknowledgments}
This work was partially supported under grants CONACyT-SEP 55310,
CONACyT 50-155I, DGAPA-UNAM IN109107 and DGAPA-UNAM IN116408.
\end{acknowledgments}

\end{document}